\documentclass[prl, twocolumn, showpacs, superscriptaddress]{revtex4-2}
\usepackage{amsmath, amssymb}
\usepackage{graphicx}
\renewcommand{\section}[1]{{\par\it #1.---}\ignorespaces}

\begin{document}
\title{First-Order Excited-State Quantum Phase Transition in the Transverse Ising Model with a Longitudinal Field}
\author{Yun-Tong Yang}
\affiliation{School of Physical Science and Technology, Lanzhou University, Lanzhou 730000, China}
\affiliation{Lanzhou Center for Theoretical Physics $\&$ Key Laboratory of Theoretical Physics of Gansu Province, Lanzhou University, Lanzhou 730000, China}
\author{Hong-Gang Luo}
\email{luohg@lzu.edu.cn}
\affiliation{School of Physical Science and Technology, Lanzhou University, Lanzhou 730000, China}
\affiliation{Lanzhou Center for Theoretical Physics $\&$ Key Laboratory of Theoretical Physics of Gansu Province, Lanzhou University, Lanzhou 730000, China}
\affiliation{Beijing Computational Science Research Center, Beijing 100084, China}

\begin{abstract}
The investigation of the first-order quantum phase transition (QPT) is far from clarity in comparison to that of the second-order or continuous QPT, in which the order parameter and associated broken symmetry can be clearly identified and at the same time the concepts of universality class and critical scaling can be characterized by critical exponents. Here we present a compared study of these two kinds of QPT in the transverse Ising model. In the absence of a longitudinal field, the ground state of the model exhibits a second-order QPT from paramagnetic phase to ferromagnetic one, which is smeared out once the longitudinal field is applied. Surprisingly, the first excited state involves a first-order QPT as the longitudinal field increases, which has not been reported in literature. Within the framework of a pattern picture we clearly identify the difference between these two kinds of QPT: for the continuous QPT only the pattern flavoring ferromagnetic phase is always dominant over the others, and on the contrary, there exist at least two competitive patterns in the first-order QPT, which is further indicated by patterns' occupancies calculated by pattern projections on the ground and first excited states wavefunctions. Our result has not only a fundamental significance in the understandings of the nature of QPTs, but also a practical interest in quantum simulations used to test the present finding.
\end{abstract}
\maketitle

\section{Introduction}
The existences of matter in nature and laboratory have vaious forms, for example, water can be liquid, gaseous, or solid, and their transitions are one of central issues in condensed matter physics and statistical physics \cite{Chaikin2000}. According to Ehrenfest's classification, the phase transitions can be categorized according to the behavior of the derivatives of energy (or free energy at finite temperature) with respect to temperature or other thermodynamical variables at transition point: if the first derivative of energy is discontinuous, then the phase transition is first-order; otherwise, the phase transition is second-order if the second derivative is discontinuous, assuming its first derivative is continuous, and the rest may be deduced by analogy. The second-order phase transition or above are usually called continuous ones, which can be described \textit{phenomenologically} by order parameter and associated broken symmetry \cite{Landau1937, Landau1980}. Furthermore, a complete theoretical description has been obtained by using renormalization group \cite{Wilson1974, Wilson1975}. In particular, the concepts of universality class and critical scaling characterized by critical exponents \cite{Chaikin2000, Wilson1974, Wilson1975}  can be used to classify the phase transitions, irrespective of what kinds of matter.  

In contrast to the situation of the continuous phase transition, the investigation of the first-order phase transition is far from clarity. The reason is that except for the latent heat realized early by the van der Waals equation of state \cite{Waals1873}, the first-order phase transition is lack of typical features characterized its critical behaviors, just like the order parameters and associated broken symmetries in the continuous ones. As a result, the concept like order parameter can only be used cautiously in the study of the first-order phase transitions \cite{Stanley1987, Kondepudi1998}. However, a robust observation is that there exists a metastable equilibrium region in the vicinity of the first-order transition point, where at least two or more possible metastable states compete each other for a long time before the first-order phase transition takes place. The formation of a new phase is usually initiated when the nucleation of the new phase commences, which has been popularly observed in liquid-gas or liquid-solid phase transitions \cite{Chaikin2000}. 

Here we present an example to exhibit a first-order quantum phase transition (QPT) involved in the simplest many-body model, namely, the transverse Ising model in the presence of a longitudinal field which makes the system non-integrable. It is well-known that the ground state of the transverse Ising model experiences a second-order QPT from paramagnetic phase at weak interaction regime to ferromagnetic one at strong interaction one in the absence of longitudinal field \cite{Sachdev2011, Coldea2010, Breunig2017}. In the previous paper \cite{Yang2022c}, we have used a pattern picture to dissect the process of this QPT, in which the patterns characterizing the ferromagnetic order becomes predominant over the others in a somehow smooth way. As a result, the QPT happens not so dramatically. On the other hand, a weak longitudinal field can induces a first-order QPT in the ferromagnetic ground state along the direction of the longitudinal field applied \cite{Yuste2018} in which we are not interested here. Increasing the longitudinal field will smear out the second-order QPT since in this case the ground state always flavors the ferromagnetic phase. In this case, we find surprisingly that interesting physics is involved in the excited states. With increasing longitudinal field the first excited state approaches rapidly to the second excited state, and they go across to lead to a first-order QPT. This finding has not been reported in literature. In the following we explore explicitly the property of the first-order QPT by using pattern picture \cite{Yang2022c} and make a compared study for the first-order QPT in the first excited state in the presence of longitudial field and continuous one in the ground state in the absence of longitudial field \cite{Yang2022c} \textit{on an equal footing}.

\section{Model and Method}
The transverse Ising model Hamiltonian with a longitudial field reads
\begin{equation}
\hat{H}' = - J'\sum_{i,\delta}\hat\sigma^z_i \hat\sigma^z_{i+\delta} - h' \sum_i \hat{\sigma}^z_i - g\sum_i \hat\sigma^x_i, \label{Ising0}
\end{equation}
where $J'$, $h'$ and $g$ are the Ising interaction between two spins denoting by Pauli matrix $\hat\sigma$ located at site $i$ and its nearest neighbors $i+\delta$ and the longitudinal and transverse fields, respectively. All these parameters are non-negative, and thus the interaction is ferromagnetic. For convenience, we take the transverse field $g$ as units of energy. This is different to the conventional notations in literature \cite{Sachdev2011} but we find that it is more convenient to use the present notations to discuss the first-order excited-state QPT. Thus Eq. (\ref{Ising0}) is rewritten as
\begin{eqnarray}
&& \hat{H}'= \frac{g}{2} \hat{H}, \; \hat{H} = \sum_{i} \hat{H}_i\label{Ising1a} \\
&& \hat{H}_i = -2\hat{\sigma}^x_i - 2h \hat{\sigma}^z_i - J\sum_\delta \left(\hat\sigma^z_i \hat\sigma^z_{i+\delta} + \hat\sigma^z_{i+\delta} \hat\sigma^z_i\right),\label{Ising1b}
\end{eqnarray}
where $J = J'/g$ and $h = h'/g$. For simplicity, we limit ourselves to the one-dimensional(1D) case, though it is straightforward to extend to the high-dimensional situations. For a chain with size $L$ under periodic boundary condition (PBC, i.e.,$\hat{\sigma}^z_{L+1} = \hat{\sigma}^z_1$), $\hat{H}$ can be reformulated as a $3L \times 3L$ matrix in a lattice operator space as follows
\begin{eqnarray}
&&\hat{H} = \left(
\begin{array}{cccccccccc}
\hat{\sigma}^x_1 & -i\hat{\sigma}^y_1& \hat{\sigma}^z_1& \hat{\sigma}^x_2& -i\hat{\sigma}^y_2& \hat{\sigma}^z_2& \cdots& \hat{\sigma}^x_L& -i\hat{\sigma}^y_L& \hat{\sigma}^z_L
\end{array}
\right)\nonumber\\
&&\times
\left(
\begin{array}{cccccccccc}
0 & h &0 &0 &0 &0 &\cdots &0 &0 &0 \\
h & 0 &- 1 &0 &0 &0 &\cdots &0 &0 &0 \\
0 & - 1 &0 &0 &0 &- J &\cdots &0 &0 &-J \\
0 & 0 &0 &0 &h &0 &\cdots &0 &0 &0 \\
0 & 0 &0 &h &0 &- 1 &\cdots &0 &0 &0 \\
0 & 0 &- J &0 &- 1 &0 &\cdots &0 &0 &0 \\
\vdots & \vdots &\vdots &\vdots &\vdots &\vdots &\ddots &\vdots &\vdots &\vdots \\
0 & 0 &0 &0 &0 &0 &\cdots &0 &h &0 \\
0 & 0 &0 &0 &0 &0 &\cdots &h &0 &- 1 \\
0 & 0 &- J &0 &0 &0 &\cdots &0 &- 1 &0 
\end{array}
\right)\nonumber\\
&&\times
\left(
\begin{array}{cccccccccc}
\hat{\sigma}^x_1 & i\hat{\sigma}^y_1 & \hat{\sigma}^z_1 & \hat{\sigma}^x_2 & i\hat{\sigma}^y_2 &\hat{\sigma}^z_2 &\cdots & \hat{\sigma}^x_L & i\hat{\sigma}^y_L & \hat{\sigma}^z_L
\end{array}
\right)^T,\label{Ising2}
\end{eqnarray}
where the identities $\hat{\sigma}^x \hat{\sigma}^y = i \hat{\sigma}^z $ and $\hat{\sigma}^y \hat{\sigma}^z = i \hat{\sigma}^x$ have been used for each site $i$ and the superscript $T$ denotes transpose of the operator vector. The matrix in Eq. (\ref{Ising2}) can be diagonalized to obtain eigenvalues and corresponding eigenfunctions $\{\lambda_n, u_n\} (n = 1, 2, \cdots, 3L)$. We call them in the following patterns marked by $\lambda_n$. With these patterns at hand, $\hat{H}$ is reformulated as
\begin{equation}
\hat{H} = \sum_{n=1}^{3L} \lambda_n \hat{A}^\dagger_n \hat{A}_n, \label{Ising3a} 
\end{equation}
where each pattern $\lambda_n$ composes of one-body operators
\begin{equation}
\hat{A}_n = \sum_{i=1}^{L}\left[u_{n,3i-2} \hat{\sigma}^x_{i}+u_{n,3i-1} (i\hat{\sigma}^y_{i}) + u_{n,3i} \hat{\sigma}^z_{i}\right]. \label{Ising3b}
\end{equation}
The validity of Eq. (\ref{Ising3a}) can be confirmed by inserting into a complete basis $|\{\sigma^z_i\}\rangle (i = 1, 2,\cdots, L)$ with $\hat{\sigma}^z_i |\{\sigma^z_i\}\rangle = \pm_i(\uparrow,\downarrow) |\{\sigma^z_i\}\rangle$. Thus one obtains a matrix $\left[\hat{A}_n\right]_{\{\sigma^z_i\},\{\sigma^z_i\}^{\prime}} = \langle\{\sigma^z_i\}|\hat{A}_n|\{\sigma^z_i\}^{\prime}\rangle$ and then Eq. (\ref{Ising3a}) can be solved by diagonalizing the matrix
\begin{equation}
\langle \{\sigma^z_i\}|\hat{H}|\{\sigma^z_i\}^{\prime}\rangle = \sum_{n=1}^{3L} \lambda_n \sum_{\{\sigma^z_i\}^{\prime\prime}} \left[\hat{A}^\dagger_n\right]_{\{\sigma^z_i\},\{\sigma^z_i\}^{\prime\prime}}\left[\hat{A}_n\right]_{\{\sigma^z_i\}^{\prime\prime},\{\sigma^z_i\}^{\prime}}.\label{Ising4}
\end{equation}
\begin{figure}[tbp]
\begin{center}
\includegraphics[width = \columnwidth]{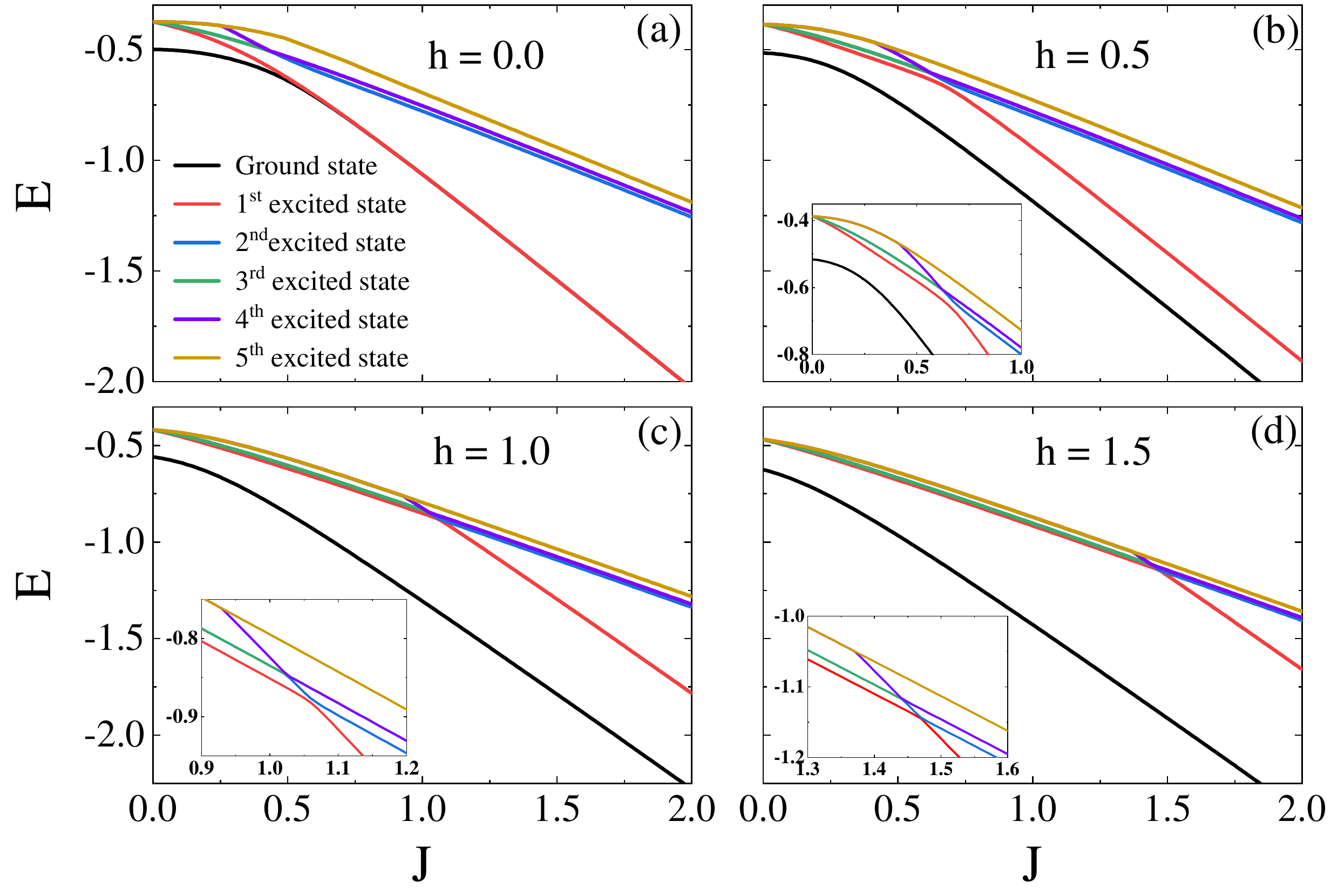}
\caption{The ground state and first five excited states energies of the 1D transverse Ising model in the absence (a) and presence (b, c, d) of longitudinal fields. The insets in (b), (c) and (d) show enlarged views of the crossovers of excited states.}\label{fig1}
\end{center}
\end{figure}
The energies of the ground state and first five excited states are provided in Fig. \ref{fig1} for several longitudinal fields $h = 0.0, 0.5, 1.0$ and $1.5$ for $L=8$ we take here and hereafter, which are complete agreement with those by numerical exact diagonalization (ED) [not shown here but those of the ground and first excited states are shown in Fig. \ref{fig4} (a1) and (b1) for comparison]. For $h = 0.0$ it is very clear that there exists a second-order QPT (see, e.g. Ref.\cite{Yang2022c}) at $J \sim 0.5$. With increasing $J$, the second-order QPT disappears since the first excited state is lifted to approach to the second and third excited states, as shown in Fig. \ref{fig1} (b) for $h = 0.5$ and at the same time, the onset of a first-order QPT is seen due to possible discontinuous change of the first excited state energy. The trend is more apparent at $h=1.0$ shown in the inset of Fig. \ref{fig1}(c) and the first-order QPT happens at $h = 1.5$, a sufficient large longitudinal field, shown in Fig. \ref{fig1}(d). In the following we study in detail this first-order QPT by using pattern picture, and compare it with the second-order QPT at $h=0.0$ \cite{Yang2022c}. 

\begin{figure}[tbp]
\begin{center}
\includegraphics[width = \columnwidth]{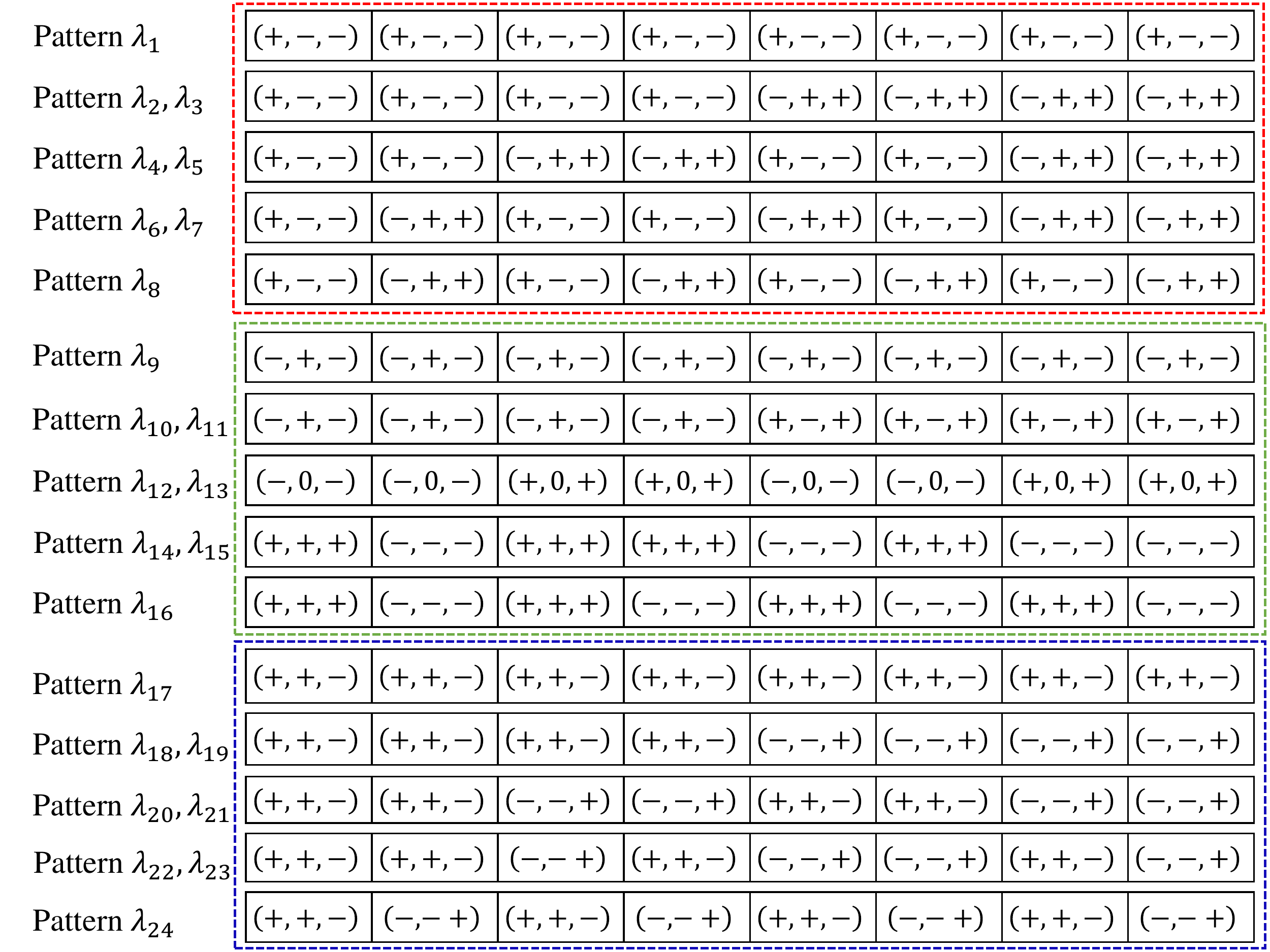}
\caption{The patterns and their relative phases obtained by the first diagonalization, marked by the single-body operators $\hat{A}_n = \sum_{i=1}^L \left[u_{n,3i-2} \hat{\sigma}^x_i+u_{n,3i-1} (i\hat{\sigma}^y_i) + u_{n,3i}\hat{\sigma}^z_i\right]$ with $(\pm,\pm,\pm)$ denoting the signs of $(u_{n,3i-2},u_{n,3i-1},u_{n,3i})$ for the 1D tranverse Ising model with $L=8$ under PBC. All patterns are divided into three groups marked by the dashed red, green, and blue frames, respectively. For each pattern, a phase factor $e^{i\pi}$ is free, not affecting the relative signs within and between patterns. }\label{fig2}
\end{center}
\end{figure}

\begin{figure}[tbp]
\begin{center}
\includegraphics[width = \columnwidth]{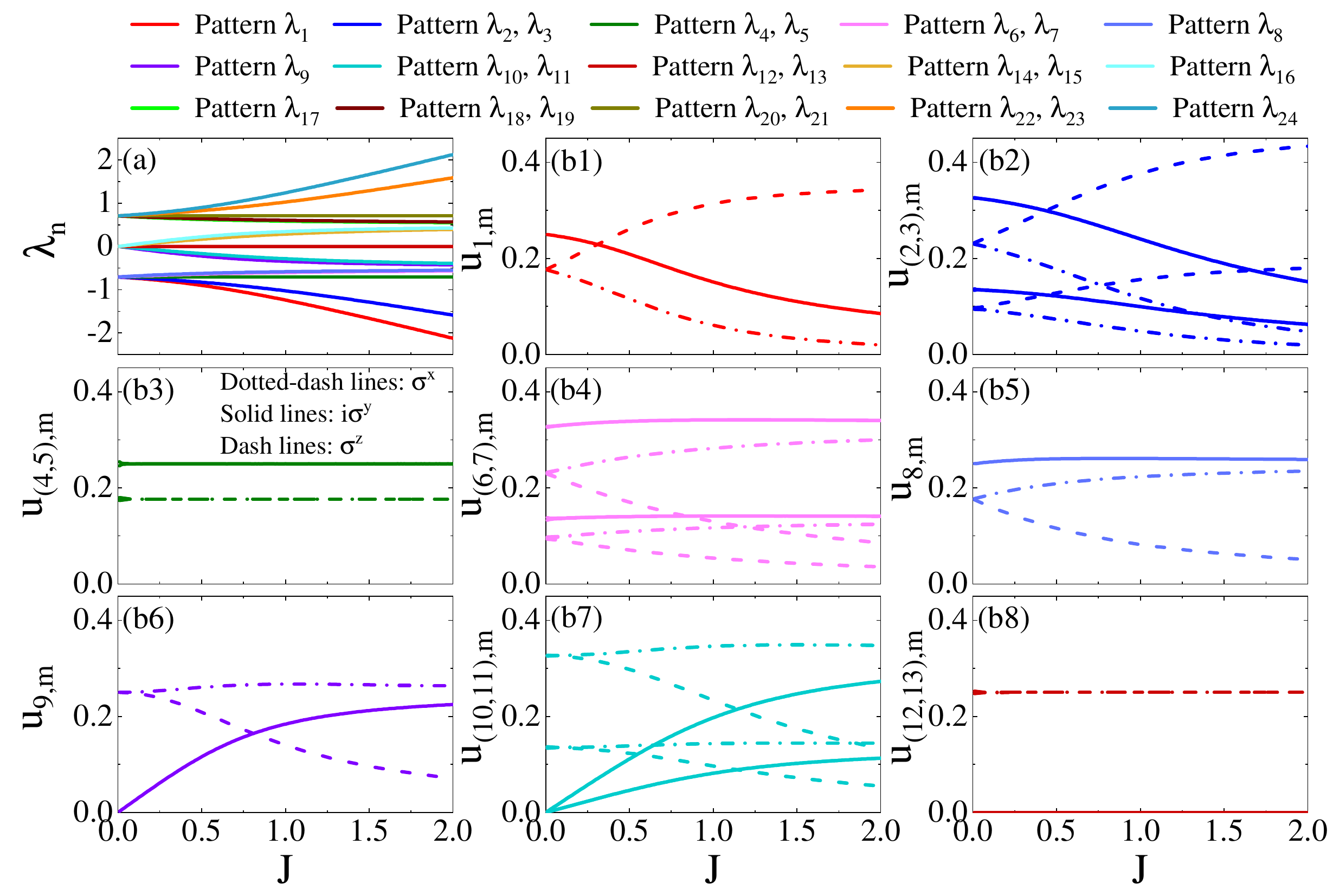}
\caption{(a) The eigenvalues of patterns and [(b1)-(b8)] their eigenfunctions as functions of $J$ and $h = 1.0$ is taken. The patterns satisfy with $\lambda_n = - \lambda_{3L-n+1}$ and $u_{n, m} = - u_{3L-n+1, m}$, where $m$ denotes the spin components. Thus the eigenfunctions from $u_{14,m}$ to $u_{24,m}$ are not shown.} \label{fig3}
\end{center}
\end{figure}
\section{Characters of Patterns} 
Figures \ref{fig2} and \ref{fig3} show the patterns and their eigenfunctions dependent of $J$. According to the behaviors of the eigenvalues [Fig. \ref{fig3} (a)], all patterns can be divided into three groups, which are degenerate for each group at $J=0$ and the degeneracy is partly lifted for increasing $J$. Their eigenfunctions are smoothly dependent or independent of $J$, showing no any singular behaviors. The signs of patterns show several characteristic features: (i) except for the patterns $\lambda_{12,13}$ with zero pattern eigenvalues, the patterns with negative eigenvalues have the opposite signs for the coefficients of $\hat{\sigma}_i^x$ and $i\hat{\sigma}_i^y$, namely, they are out-of-phase; (ii) oppositely, they are in-phase for the patterns with positive eigenvalues; (iii) the distinguish between different patterns in each group is the signs of the coefficients of $\hat{\sigma}_i^z$ and their orders, which correspond to different domains or kinks contained in the patterns; (iv) the patterns $\lambda_1$, $\lambda_9$ and $\lambda_{17}$ are specially important since their eigenvalues are the lowest one in their own groups, and there is a common character, namely, the coefficients of $\hat{\sigma}_i^z$ are in-phase, which are possible candidates of metastable states; (v) in each group the patterns with two domains or kinks are also potential candidates of metastable states, which are important to the excited state QPT we study. An additional remark on the chain size is in order. The above observations are independent of the chain size. More larger the chain size is, more patterns are, but the above characters of patterns remain valid, only effect is to increase the number of patterns with more domains or kinks. This is the reason that the physics of QPT can be essentially captured, even for small chain size, as discussed here.

\begin{figure}[tbp]
\begin{center}
\includegraphics[width = \columnwidth]{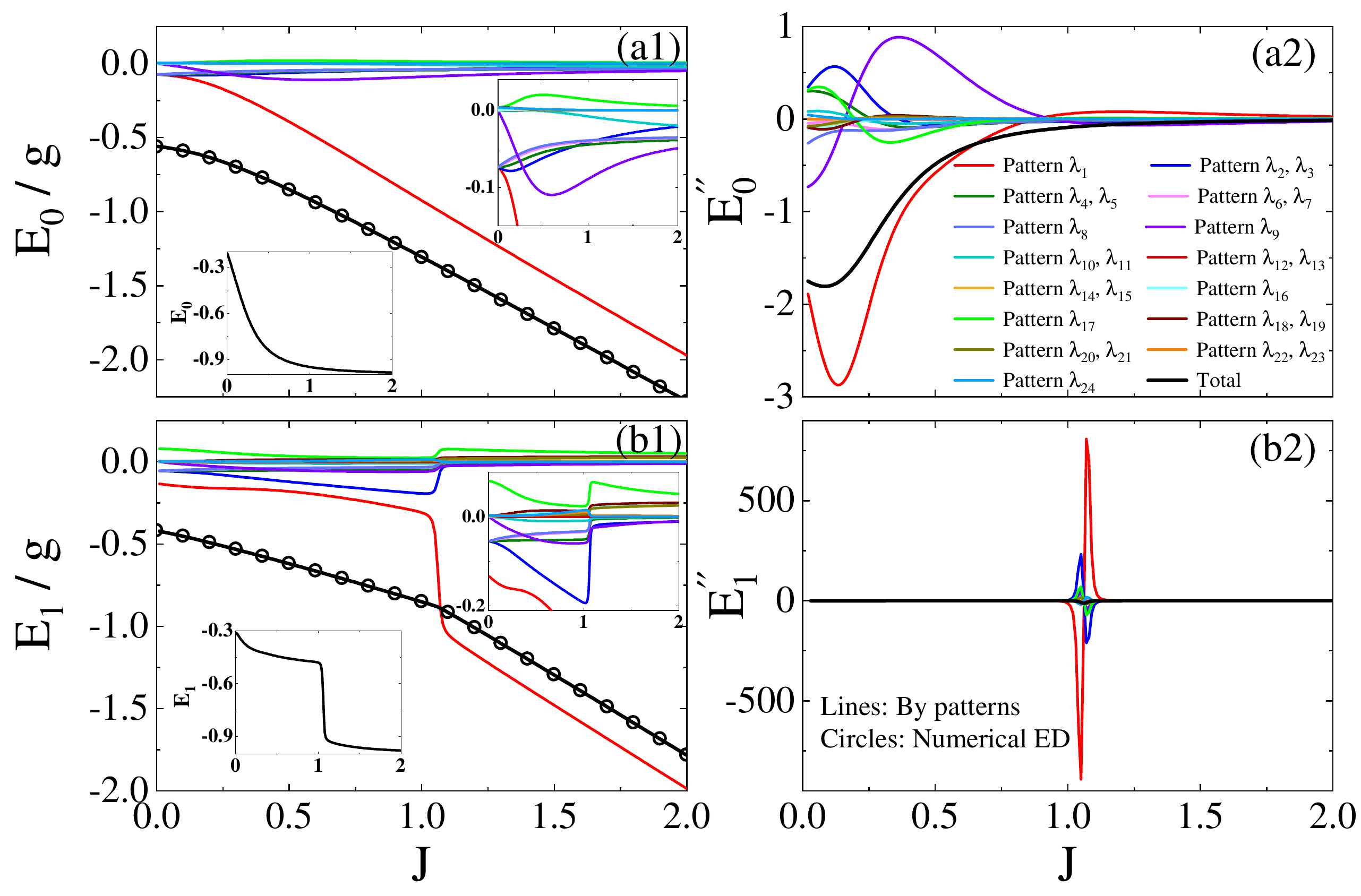}
\caption{(a1) $\&$ (b1) The ground state and first excited state energies as functions of $J$ [thick black lines (calculated by patterns) and circles (numerical ED)]. The left-lower insets in (a1) and (b1) denote the first derivatives of them. The right-upper insets in (a1) and (b1) are enlarged views of the energy components of patterns. (a2) $\&$ (b2) The second derivatives of the energy components of patterns. Here $h=1.0$ is taken.}\label{fig4}
\end{center}
\end{figure}
\section{First-Order Excited-State QPT} 
Figure \ref{fig4} (a1) $\&$ (b1) present energies of the ground and first excited states as functions of the interaction $J$, respectively, as shown by thick black solid lines. The results of numerical ED (circles) are also presented for comparison, which confirms the validity of the pattern formulation, as also mentioned above. As shown in Fig. \ref{fig1} (b)-(d), the ground state is always dominant over the first excited state, and in this case the system is always ferromagnetic. Therefore, there is no QPT happening for the ground state, as also seen from the smooth first derivative shown in the inset and second derivative in Fig. \ref{fig4} (a2). 

Turning to the first excited state, the situation changes dramatically. An obvious discontinuity of the first excited state energy is seen near $J \sim h = 1.0$, as shown in the inset of Fig. \ref{fig4} (b1). This first-order QPT is more apparent by checking the contributions of pattern components to the first excited state energy, specially that of the pattern $\lambda_1$. Other components are enlargedly shown in the right-upper inset of Fig. \ref{fig4} (b1). The energy of the pattern $\lambda_1$ has a sudden drop at $J \sim h$, even lower than the energy of the first excited state. At the same time, other patterns like $\lambda_{2,3}$, $\lambda_{9}$, and $\lambda_{17}$ also have dramatic changes. They indeed form metastable states, which have a local minimum. In particular, the patterns $\lambda_{2,3}$ even directly compete with the pattern $\lambda_1$, they have comparable energy contributions to the first excited state at $J \leq h$. Another important observation is the response of the patterns with positive eigenvalues, which are excited by the first-order QPT and their suppressions by increasing the interaction strength $J$ are slow. 

\begin{figure}[tbp]
\begin{center}
\includegraphics[width = \columnwidth]{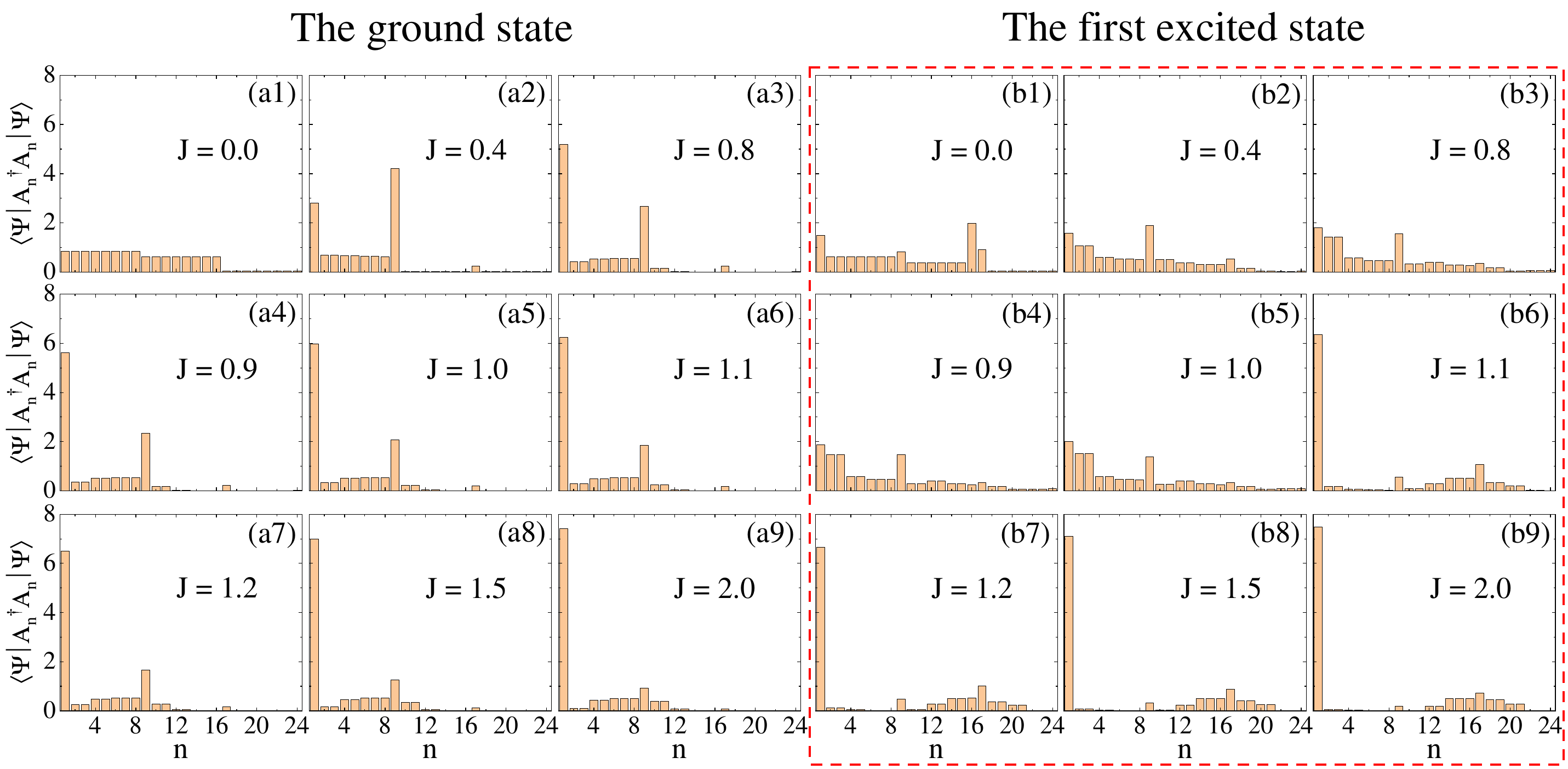}
\caption{Histograms of patterns' occupancy of the ground state [(a1)-(a9)] and the first excited state [(b1)-(b9), grouped by dashed red frame] of the system with $h=1$ for different Ising interactions $J = 0.0, 0.4, 0.8, 0.9, 1.0, 1.1, 1.2, 1.5$, and $2.0$, which correspond successively from [(a1),(b1)] to [(a9),(b9)], respectively.}\label{fig5}
\end{center}
\end{figure}

The above observations can be further confirmed by checking the histograms of the patterns' occupancies, as shown in Fig. \ref{fig5} for $h = 1.0$. The patterns' occupancies calculated by $\langle\Psi|\hat{A}^\dagger_n \hat{A}_n|\Psi\rangle$ where $|\Psi\rangle$ is the ground state for (a1)-(a9) and the first excited state for (b1)-(b9) (marked by dashed red frame) in Fig. \ref{fig5}. For the ground state, it is seen that the pattern occupancies (heights of the patterns) vary with increasing $J$ in a quite smooth way, more smoother than the ground state at $h=0.0$ (a second-order QPT exists \cite{Yang2022c}), so that the QPT smears out, although the occupancies behave in a similar way. For example, that of the pattern $\lambda_1$ increases gradually and finally it dominates over the other patterns; that of the pattern $\lambda_9$ has a rapid increase for small $J$, even beyond that of the pattern $\lambda_1$, but its height is suppressed gradually, even if its contribution is still visible. A similar behavior is also observed for the pattern $\lambda_{17}$, the only difference is that its contribution is negligible. 

The situation is quite different in the first excited state. With increasing $J$, the heights of occupancies of the patterns $\lambda_1$ and $\lambda_{2,3}$ increase at a similar rate up to $J = 1.0$, and the occupancy of the pattern $\lambda_9$ is heighest at $J = 0.4$, but it is suppressed gradually up to $J = 1.0$. Increasing the interaction strength up to $J = 1.1$, a slightly higher than $h = 1.0$, the pattern $\lambda_1$ becomes rapidly dominant, the patterns $\lambda_{2,3}$ become negligible, which means a first-order QPT takes place. A similar loss of height also happens in the pattern $\lambda_9$, even if its suppression is not so obvious in comparison to that of the patterns $\lambda_{2,3}$. More surprisingly, an opposite thing happens in the pattern $\lambda_{17}$ and its near patterns. Their heights have an obvious increase, and then are suppressed gradually with increasing $J$. These observations are in agreenment with those from the varyings of energy contributions of the patterns. It is useful to compare the histograms of patterns' occupancies in two kinds of QPTs (the second-order one at $h = 0.0$ in the ground state \cite{Yang2022c} and the first-order one at $h = 1.0$ in the first excited state). In the second-order QPT only the pattern, namely, pattern $\lambda_{1}$, flavoring ferromagnetic phase is always dominant over others. However, in the first-order one there exist at least two competitive patterns such as pattern $\lambda_{1}, \lambda_{2,3}$ and $\lambda_{9}$ which form metastable states. This is the main difference of these two kinds QPTs. 

In addition, it is also of interest to compare the pattern occupancies between the ground and first excited states shown in \ref{fig5}. A common character is that the pattern $\lambda_1$ dominates over the others for large $J$, which both flavor the ferromagnetic phase. The difference is the occupancies of the other patterns due to quantum fluctuations. In the ground state, the other occupancies mainly distribute near the pattern $\lambda_9$ and in the first excited state these occupancies mainly locate near the pattern $\lambda_{17}$. Thus these two cases are complementary, which may be useful in quantum annealing \cite{Kadowaki1998, Das2008, Johnson2011, Grass2019}, which involves dynamics of the model we study. This is obviously beyond the scope of the present work and is left for the future study. 

\section{Summary and Discussion}
The well-known second-order QPT in the transverse Ising model is smeared out once a longitudinal field is applied. Instead, a first-order QPT is found in the first excited state, which has not been reported in literature. We provide a pattern picture obtained by two successive diagonalizations to study these two QPTs on an equal footing. While the second-order QPT in the absence of longitudinal field has been studied in details in Ref. \cite{Yang2022c}, here we mainly focus on the case with finte longitudinal field. 

After the energy contributions of different patterns to the ground and first excited states have been clearly analyzed, how to smear out the second-order QPT and how to form the first-order one are further identified. In particular, the comparison between the histograms of patterns' occupancies of the first excited state in the presence of the longitudinal field and those of the ground state in the absence of the longitudinal field characterizes the different process of the first- and second-order QPTs: in the second-order one \cite{Yang2022c}, the patterns flavoring the ferromagnetic phase are dominant over others and the phase transition is somehow not so dramatic; on the contrary, in the first-order case we study here, there are at least two competitive patterns, and as a result, the first-order QPT is so dramatic.

The first-excited QPT obtained in the present work could be tested by current quantum simulation platforms \cite{Georgescu2014} ranging from trapped ions \cite{Friedenauer2008, Monroe2021}, quantum superconducting circuits \cite{You2005, Johnson2011} to optical lattice \cite{Greiner2008}, and so on. The non-integrability of the system in the presence of a longitudinal field further richens the phase transition physics involved such a paradigmatic model.

\section{Acknowledgments}
The work is partly supported by the National Key Research and Development Program of China (Grant No. 2022YFA1402704) and the programs for NSFC of China (Grant No. 11834005, Grant No. 12247101).




%

\end{document}